\documentclass{article}
\usepackage{spconf,amsmath,graphicx, booktabs}
\usepackage{multirow}
\usepackage{amssymb}
\usepackage{hyperref}
\usepackage[T1]{fontenc}
\usepackage[utf8]{inputenc}

\title{PFluxTTS: Hybrid Flow-Matching TTS with Robust Cross-Lingual Voice Cloning and Inference-Time Model Fusion}

\name{
  \textit{Vikentii Pankov\textsuperscript{1}, Artem Gribul\textsuperscript{1}, Oktai Tatanov\textsuperscript{1}, Vladislav Proskurov\textsuperscript{1},}\\
  \textit{Yuliya Korotkova\textsuperscript{2}, Darima Mylzenova\textsuperscript{3}, Dmitrii Vypirailenko\textsuperscript{1}}
  \thanks{© 2026 IEEE. Personal use of this material is permitted. Permission from IEEE must be obtained for all other uses, in any current or future media, including reprinting/republishing this material for advertising or promotional purposes, creating new collective works, for resale or redistribution to servers or lists, or reuse of any copyrighted component of this work in other works.}
}
\address{
  \textsuperscript{1}Rask AI, USA \quad
  \textsuperscript{2}École Polytechnique, France \quad
  \textsuperscript{3}TBC Bank, Uzbekistan
}

\begin{document}
\ninept
\maketitle

\begin{abstract}
We present PFluxTTS, a hybrid text-to-speech system addressing three gaps in flow-matching TTS: the stability-naturalness trade-off, weak cross-lingual voice cloning, and limited audio quality from low-rate mel features. Our contributions are: (1) a dual-decoder design combining duration-guided and alignment-free models through inference-time vector-field fusion; (2) robust cloning using a sequence of speech-prompt embeddings in a FLUX-based decoder, preserving speaker traits across languages without prompt transcripts; and (3) a modified PeriodWave vocoder with super-resolution to 48\,kHz. On cross-lingual in-the-wild data, PFluxTTS clearly outperforms F5-TTS, FishSpeech, and SparkTTS, matches ChatterBox in naturalness (MOS 4.11) while achieving 23\% lower WER (6.9\% vs. 9.0\%), and surpasses ElevenLabs in speaker similarity ($+0.32$ SMOS). The system remains robust in challenging scenarios where most open-source models fail, while requiring only short reference audio and no extra training. Audio demos are available at \url{https://braskai.github.io/pfluxtts/}.
\end{abstract}

\section{Introduction}
\label{sec:intro}

Recent progress in text-to-speech (TTS) has been driven by flow-matching (FM) architectures, which learn vector fields for fast, high-fidelity synthesis via ODE-based inference~\cite{mehta2023matcha,blackforest2025flux}. 
Despite these advances, FM-TTS systems still face three core gaps: 
\emph{alignment} --- stability and control in duration-guided models versus fluency and naturalness in alignment-free models remain a trade-off; 
\emph{cross-lingual voice cloning} --- fixed speaker vectors discard time-varying timbre, while more expressive prompt conditioning can destabilize alignment-free models, leaving robust use of long prompts largely unexplored; 
\emph{vocoder mismatch} --- full-band $48$\,kHz reconstruction from low-rate mel features is underexplored in TTS models.

\noindent\textit{Duration-Guided vs. Alignment-Free Models.} Many flow-matching TTS systems (e.g., Matcha-TTS~\cite{mehta2023matcha}, Voicebox~\cite{voicebox}, P-Flow~\cite{kim2023pflow}) rely on explicit duration predictors. This produces overly smoothed, low-variance durations and limits prosodic naturalness. While VITS~\cite{kim2021vits} introduced a flow-based stochastic duration module, stability issues have limited practical adoption. E2-TTS~\cite{eskimez2024e2tts} removes the duration predictor entirely, but suffers from stability issues and slower convergence. F5-TTS~\cite{chen2024f5} partially mitigates these issues via ConvNeXt-refined text representations enabling implicit text–speech alignment, yet stability problems (e.g., word skipping) persist, especially in challenging conditions such as noisy cross-lingual voice cloning.

An alternative is autoregressive (AR) modeling ~\cite{chen2024valle2}, which achieves strong naturalness but suffers from slow inference and limited robustness. Several designs target these limitations: Fish-Speech~\cite{fish-speech-v1.4} improves codebook efficiency with Grouped Finite Scalar Vector Quantization and a fast–slow dual AR generator; and Seed-TTS~\cite{anastassiou2024seedtts} explores both AR over discrete tokens and a non-autoregressive (NAR) variant that predicts total utterance duration and then allocates per-phoneme lengths --- promising, though it was evaluated for content and speaking-rate editing only. To bridge models without explicit alignment and those relying on predefined alignments, MegaTTS-3~\cite{jiang2025megatts3} introduces sparse alignment that guides a latent DiT using boundaries from a duration predictor, easing alignment without over-constraining the search space.

\noindent\textit{Voice Cloning. } Most zero-shot voice cloning systems rely on fixed-dimensional speaker embeddings. These systems often struggle with cross-lingual prompts and speaker identity preservation. In particular, compressing speaker information into a single embedding discards fine-grained, time-varying attributes --- especially in longer utterances. This issue is addressed in XTTS~\cite{casanova2024xtts}, which conditions on a set of Perceiver-Resampler prompt embeddings rather than a single embedding for zero-shot AR TTS. To benefit from longer prompts and extract fine-grained timbre information, MegaTTS-2~\cite{jiang2023megatts2} uses multiple audio prompts from the target speaker that differ from the target mel-spectrogram. To keep the timbre stable while controlling prosodic style separately MegaTTS-2~\cite{jiang2023megatts2} and NaturalSpeech 3~\cite{ju2024naturalspeech3} additionally disentangle timbre, prosody, and acoustic content. Models like Voicebox~\cite{voicebox} or codec-based AR LMs~\cite{chen2024valle2} achieve strong identity cloning via audio infilling or continuation.

\noindent\textit{Vocoder Enhancements.} 
Conditional Flow Matching (CFM) has also been applied to vocoder design. PeriodWave~\cite{lee2025periodwave, lee2024periodwave_turbo} employs CFM for waveform generation and achieves state-of-the-art fidelity.
Another line of work explores neural vocoders with super-resolution, such as NVSR~\cite{liu2022neural}, based on a ResUNet architecture, and the diffusion-based NU-Wave2~\cite{han2022nuwave2}; both operate at multiple input resolutions and enhance synthesis from lower sample rates. 

To summarize, we propose PFluxTTS, a hybrid FM-TTS system that addresses the challenges related to alignment, voice cloning, and vocoder quality described above. Our contributions are as follows:
\begin{enumerate}
    \item We introduce a dual-decoder architecture that combines an alignment-free (AF) model with a duration-guided (DG) model through inference-time vector-field fusion, combining the stability of explicit durations with the naturalness and fluency of alignment-free decoding. 
    \item We design a voice cloning strategy that employs a sequence of speech-prompt embeddings within a FLUX-based architecture, which is robust to long, cross-lingual prompts and does not require prompt text.
    \item We integrate a PeriodWave-based vocoder with prompt-based super-resolution, enabling $48$\,kHz waveform reconstruction from low-rate mel features. 
\end{enumerate}

Comprehensive experiments on challenging cross-lingual and in-the-wild conditions show that PFluxTTS achieves higher intelligibility and speaker similarity than state-of-the-art baselines.

Section~2 details the proposed method, including the architecture,  inference-time fusion, voice cloning, and vocoder. Section~3 presents experimental setup and results, and Section~4 concludes.

\section{Proposed method}

\subsection{Overall architecture}
\label{sec:overview}

We synthesize a mel-spectrogram $\hat{\mathbf{m}}\!\in\!\mathbb{R}^{F\times T}$ from phonemes $\mathbf{p}$ and an acoustic prompt $\mathbf{s}$. We utilize two TTS models trained independently with no weight sharing between them: a \emph{duration-guided} model (DG) and an \emph{alignment-free} model (AF). 
Each model has its own text and prompt encoders that produce text features $c_{\text{text}}^{\mathrm{DG/AF}}$ and prompt features $c_{\text{sp}}^{\mathrm{DG/AF}}$; the TextEncoder additionally conditions on a language ID and a pretrained ECAPA-TDNN speaker embedding via AdaLN.

\begin{figure*}[t]
    \centering
    \includegraphics[width=1.0\linewidth]{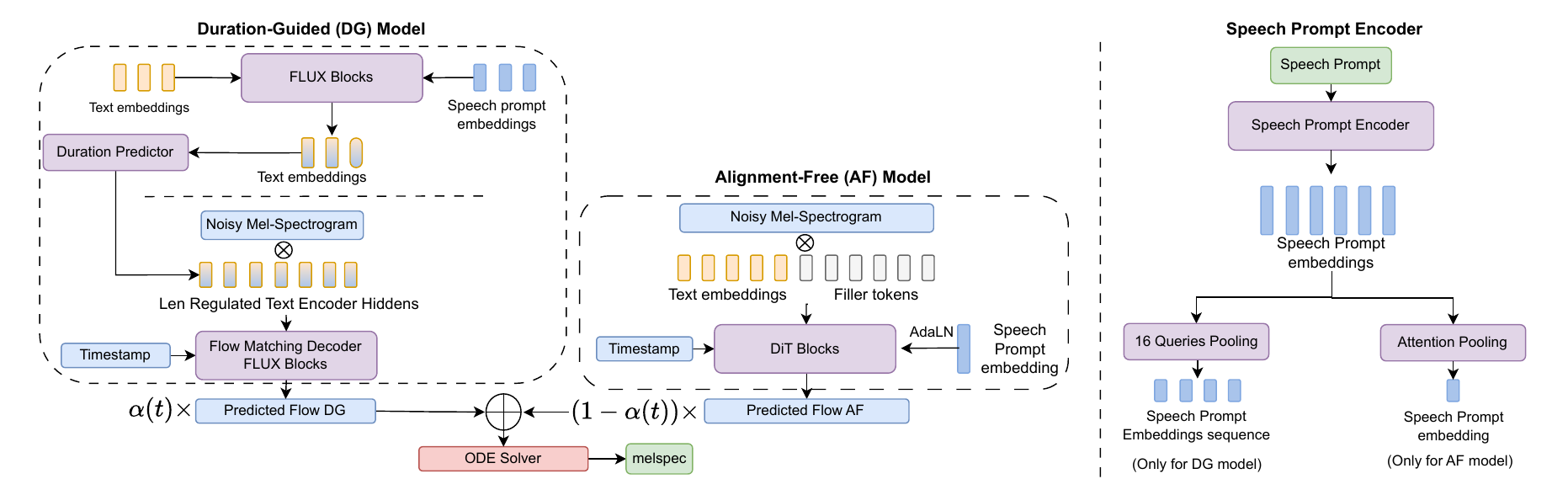}
    \caption{Architecture of PFluxTTS. Duration-Guided and Alignment-Free models are mixed with schedule $\alpha(t)$ during inference. On the right, Speech Prompt Encoder is shown, which outputs either an embedding sequence for the DG model or fixed embedding for the AF model.}
    \label{fig:arch}
\end{figure*}

At inference, the DG and AF vector fields are fused within a single ODE integration (Sec.~\ref{sec:prelim}) to obtain $\hat{\mathbf{m}}$, which is converted to 48\,kHz speech by a PeriodWave-based super-resolution vocoder. 
The overall architecture is illustrated in Fig.~\ref{fig:arch}.

\noindent\textit{DG (duration-guided) path.}
We follow FLUX \cite{blackforest2025flux} and propose a decoder with eight \emph{DoubleStream} blocks followed by 16 \emph{SingleStream} blocks. 
In DoubleStream, prompt and content tokens use separate parameters and interact via self-attention over the concatenated sequence; 
SingleStream jointly refines the merged representation, after which we retain content tokens only. Additionally, we insert a FLUX block with the same architecture as the flow-matching decoder before the length regulator and the CFM decoder, so that the text embeddings are enriched with prompt information at an early stage.

\noindent\textit{AF (alignment-free) path.}
Following F5-TTS \cite{chen2024f5}, we use a DiT-style conditional decoder and expand the phoneme sequence with learned \emph{filler} tokens to match the acoustic length $T$ and predict the mel with CFM \emph{without} a duration module. 
Alignment is learned implicitly; at inference we reuse the $T$ predicted by the duration predictor from the DG model so both paths operate on the same $(F,T)$ grid.

\subsection{Voice cloning}
\label{sec:voicecloning}
We condition the system on an acoustic prompt to preserve speaker identity across languages and recording conditions. 
Each path employs its own \emph{SpeechPromptEncoder} matched to the decoder design. 

In the \textbf{DG} path, an 8-layer Transformer encodes the prompt mel-spectrogram. 
The \emph{SpeechPromptEncoder} reduces this variable-length sequence to a set of $K=16$ embeddings via learnable query pooling, each matching the decoder’s embedding dimension. 
These tokens are attended jointly with content tokens in the FLUX decoder. 
Although we initially experimented with alternative downsamplers such as TDNNs (which preserve frame-level structure), the final system adopts query pooling ($K=16$) for its simplicity and stability.

In the \textbf{AF} path, the same backbone followed by self-attention pooling produces a \emph{fixed} 1024-d prompt embedding $c_{\text{sp}}^{\mathrm{AF,emb}}$. 
In our experiments, sequence-level prompt conditioning in AF led to frequent word skipping, which occurred less frequently under fusion but still persisted. 
Therefore, we retain a sequence of speech-prompt embeddings in DG but adopt the fixed embedding in AF for stability.

We also condition the text encoders with a pretrained ECAPA-TDNN \cite{desplanques20_interspeech} embedding (512-d) via AdaLN; this accelerated convergence and yielded a small improvement in cloning quality. 

During training, we sample a random $1$–$6$\,s crop from the reference audio as the acoustic prompt; the same span is masked in the target mel to avoid content leakage, following F5-TTS~\cite{chen2024f5}.

\subsection{Flow-matching inference mixing}
\label{sec:prelim}

\noindent\textit{Preliminaries.}  We adopt optimal transport conditional flow matching for mel generation~\cite{lipman2023flow}. 
Given $x_1\!\sim\!q$ (ground-truth mel) and $x_0\!\sim\!\mathcal{N}(0,I)$, we use the linear path and target field
$$
x_t=(1-(1-\sigma)t)\,x_0+t\,x_1,\qquad 
u_t(x_0,x_1)=x_1-(1-\sigma)x_0,
$$
where $\sigma = 0.01$ is the minimum noise level, and train a neural vector field $v_\theta(t,x)$ by
$$
\mathcal{L}_{\mathrm{CFM}}(\theta)=\mathbb{E}_{t\sim\mathcal{U}[0,1],\,x_1\sim q,\,x_0\sim \mathcal{N}(0,I)}
\big\|\,v_\theta(t,x_t)-u_t(x_0,x_1)\big\|_2^2 .
$$
In our system, $v_\theta$ is conditioned on text and prompt features $(c_{\text{text}},c_{\text{sp}})$; for DG we use a \emph{sequence} $c_{\text{sp}}^{\mathrm{DG,seq}}$, while for AF we use a \emph{fixed} embedding $c_{\text{sp}}^{\mathrm{AF,emb}}$ (Sec.~\ref{sec:voicecloning}). 
We employ joint classifier-free guidance (CFG) that nulls both text and prompt:
\[
v_\theta^{\mathrm{cfg}}(t,x)=v_\theta(t,x\mid c_{\text{text}},c_{\text{sp}})+
\gamma\big[v_\theta(t,x\mid c_{\text{text}},c_{\text{sp}})-v_\theta(t,x\mid \varnothing)\big],
\]
with guidance strength $\gamma$. We apply conditional dropout ($p\!=\!0.1$) during training, independently zeroing $c_{\text{text}}$ or $c_{\text{sp}}$ pathways (or both).

\noindent\textit {Mixing. } We combine two independently trained TTS models at inference by fusing their vector fields. Let $v_{\theta}^{\mathrm{DG}}(t,x)$ denote the duration-guided field and $v_{\phi}^{\mathrm{AF}}(t,x)$ the alignment-free field. At each solver step, we replace a single-field update with
\[
\hat{v}(t,x_t) = \alpha(t)\, v_{\theta}^{\mathrm{DG,cfg}}(t,x_t) + \big(1-\alpha(t)\big)\, v_{\phi}^{\mathrm{AF,cfg}}(t,x_t)
\]
and integrate $\tfrac{dx_t}{dt}=\hat{v}(t,x_t)$ over $t\!\in\![0,1]$ with a midpoint ODE solver. 
In our experiments, $\alpha(t)$ is piecewise-constant: we set $\alpha(t)=\alpha$ for the first $N_1$ of $N$ solver steps and $\alpha(t)=0$ for the remaining steps, so generation finishes under the AF field only. This lets the DG field stabilize alignment early while preserving the fluency benefits of the AF model in the final steps. We analyze the effect of different $\alpha$ values in Sec.~\ref{sec:results}. When $\alpha = 0$, CER increases, largely due to word skipping, especially under noisy prompts. In contrast, $\alpha = 1$ yields more stable alignments but can over-regularize timing and reduce naturalness, as reflected by the subjective CMOS results (Sec.~\ref{sec:results}).

To ensure temporal alignment, the AF model uses the total duration $T$ predicted by the DG model, so both fields operate on the same mel shape $(F,T)$. We also normalize mel targets with the same mean/std for both models.

\subsection{PeriodWave vocoder with super-resolution}
As the waveform decoder, we use PeriodWave~\cite{lee2024periodwave_turbo}, retrained from scratch on multilingual 48\,kHz audio, with two modifications.

\noindent\textit{(i) Time-downsampled conditioning \& 48\,kHz synthesis.}
Our TTS decoders emit low frame-rate mel features computed at 24\,kHz with hop 512 (whereas many TTS systems and pretrained vocoders assume hop 256 at 24\,kHz). 
We therefore retrain PeriodWave to consume this low-rate conditioning and insert an additional upsampling block (\emph{UBlock} in~\cite{lee2024periodwave_turbo}) and a downsampling block (\emph{DBlock}) with stride 4 into the \emph{Period-Aware Estimator}, so that the model synthesizes full-band 48\,kHz audio and reconstructs high-frequency details missing in input mel.

\noindent\textit{(ii) Prompt-aware conditioning.}
A global prompt embedding $c_{\text{sp}}$ is extracted from \emph{48\,kHz} audio using ConvNeXt V2-P~\cite{woo2023convnext} encoder with attention pooling and embedding size $192$. 
After linear projection, it is added to the activations of the \emph{PeriodWave Mel Encoder}, providing complementary speaker information that compensates for high-frequency detail lost in the 24\,kHz / hop-512 mel representation. 
During training, $c_{\text{sp}}$ is computed from a short prompt segment disjoint from the target region, preventing content leakage.

\section{Experiments}

We evaluate the proposed system against pretrained open-source TTS baselines using objective and subjective metrics, with ablations of key components. Complementary datasets are used: VoxLingua-dev for objective metrics across 33 source languages, mTEDx for dubbing-style subjective evaluation across four languages, and VCTK for vocoder benchmarking.

\label{sec:format}

\subsection{Experiment setup}

\noindent\textit{Architecture details.}  
The text encoder of both DG and AF models is an 8-layer Transformer with hidden size $d=768$, taking phoneme inputs extracted by \texttt{espeak-ng}. We employ RoPE for positional encoding. The speech prompt encoder is an 8-layer Transformer backbone with the same hidden size. During training, we use a separately pretrained monotonic aligner closely following One-TTS-Alignment~\cite{badlani2022one}. For inference, durations are predicted by a lightweight 2-layer CNN with kernel size 3.  

The DG flow-matching decoder consists of eight Double-Stream and 16 Single-Stream layers with hidden size $d=768$ and head dimension 48. The AF decoder has 16 layers with $d=1024$ and head dimension 128. Inference is performed with 30 ODE steps in FP16 precision, using a fusion schedule of $\alpha=0.7$ for the first 20 steps and $\alpha=0$ for the remainder, with CFG strength set to $1.34$.

\noindent\textit{Training data.}
We collect multilingual conversational audio (\texttt{en, es, de, fr, it, pt, ru}) from Yodas \cite{li2023yodas} and other sources with a unified pipeline: pyannote diarization \cite{Bredin23}; language ID with VoxLingua107 ECAPA\textendash TDNN \cite{valk2021slt}; transcription with Whisper-tiny \cite{radford2022whisper}; and boundary refinement via SeamlessM4T UnitY2 forced alignment \cite{seamless2023}. Segments are sentence-split, boundaries refined by Silero VAD \footnote{\url{https://github.com/snakers4/silero-vad}}, and adjacent sentences merged using pause-based heuristics. We normalize leading/trailing silences to 50--200\,ms with linear fades, then apply multi-stage QA: sample rate $>$24\,kHz, re-run LID, extraneous-sound detection with CED Base \cite{dinkel2023cedconsistentensembledistillation}, and a single-speaker check via re-diarization. All filter-passed segments are re-transcribed with Whisper large-v2 \cite{radford2022whisper}, with forced alignment and hallucination filtering. Overall, $\sim$28\% of candidates pass (about 50k hours in total) and are used for training. The vocoder is trained on 3.4k hours of clean 48\,kHz data, following ~\cite{lee2025periodwave,lee2024periodwave_turbo}.

\noindent\textit{Training details.} We trained our models on 4$\times$NVIDIA A100 GPUs for 1.5M iterations with a global batch size of 128. 
Optimization used AdamW with an initial learning rate of $1\mathrm{e}{-4}$, linearly decayed to $1\mathrm{e}{-6}$ over the last 100k steps. 
To stabilize training, we applied logits softcapping with threshold $70$, following Gemma~\cite{Riviere2024Gemma2I}, which mitigated gradient instabilities. We also set max gradient clip to $5$.

\noindent\textit{Evaluation details.}  
For subjective evaluation, we sampled 40 utterances (one per source) from the \texttt{mTEDx} test subset~\cite{salesky21_interspeech}, synthesizing English speech conditioned on prompts in \texttt{es}, \texttt{fr}, \texttt{it}, and \texttt{pt}. Samples were evaluated on the Prolific platform using the AI tasker pool, restricted to native English speakers. Each system output was rated on a 1–5 scale for two criteria --- MOS naturalness and similarity mean opinion score (SMOS) --- by at least seven independent annotators, with each annotator evaluating ten samples. All systems were anonymized and randomized. To ensure reliability, we included degraded TTS anchors for both criteria and excluded annotators who consistently assigned these anchors unrealistically high scores.

For objective metrics, we use VoxLingua-dev~\cite{valk2021slt} with up to 15 samples from 33 languages (397 total), paired with random English texts from SOMOS~\cite{maniati_2022_7378801} and from ELLA-V-hard~\cite{song2025ella} for ablations.

\noindent\textit{Cross-lingual setup.}
In all tests, the synthesized speech is in English, while acoustic prompts are in other languages. Raters judged speaker similarity (SMOS) with respect to the non-English prompt voice; thus, SMOS and SPK-SIM quantify cross-lingual voice cloning.

Note that most prior TTS systems were evaluated only on clean datasets and in monolingual setups (e.g., LibriSpeech, VCTK). In contrast, we specifically design our experiments for cross-lingual, in-the-wild samples, including conversational speech, to demonstrate robustness for real-world AI dubbing applications.

\noindent\textit{Baselines.} 
We compare our system against open-source state-of-the-art baselines and the commercial ElevenLabs Multilingual v2 model, using official inference code with default configurations for ChatterBox\footnote{\url{https://github.com/resemble-ai/chatterbox}}, FishSpeech S1-mini\footnote{\url{https://github.com/fishaudio/fish-speech}}, F5-TTS\footnote{\url{https://github.com/SWivid/F5-TTS}}, and SparkTTS\footnote{\url{https://github.com/SparkAudio/Spark-TTS}}. As reported by their authors, all baselines were trained on larger multilingual datasets; we restrict evaluation to English-only synthesis, where systems are robust while non-English quality varies substantially.

\subsection{Results}
\label{sec:results}

\noindent\textit{Comparison with baselines. } The results of subjective evaluation are presented in Table \ref{comp}. The experiment shows that the proposed system is statistically better than FishSpeech in Naturalness MOS, and than ElevenLabs in SMOS (paired t-test, $p<0.05$), while performing similarly to ChatterBox in both Naturalness and SMOS.

\begin{table}[ht]
\centering
\caption{Subjective MOS ($\pm$95\% CI) on \texttt{mTEDx-test} for naturalness and speaker similarity.}
\begin{tabular}{lcc}
\toprule
\textbf{System} & \textbf{Nat. MOS} & \textbf{SMOS} \\
\midrule
PFluxTTS (ours)   & \bf{4.11} $\pm$ 0.14 & 3.51 $\pm$ 0.17 \\
ChatterBox    & 4.05 $\pm$ 0.11 & \bf{3.63} $\pm$ 0.15 \\
ElevenLabs    & 4.01 $\pm$ 0.12 & 3.19 $\pm$ 0.16 \\
FishSpeech    & 3.58 $\pm$ 0.13 & 3.60 $\pm$ 0.13 \\
\bottomrule
\end{tabular}
\label{comp}
\end{table}

We present objective metrics in Table \ref{tab:objective}. The proposed system was found to be statistically better than all compared systems across WER, CER, and SPK-SIM (Wilcoxon signed-rank test, Holm-adjusted $p<0.05$). FishSpeech, SparkTTS, and F5-TTS show low intelligibility, frequently skipping words under noisy cross-lingual speech prompts, while the proposed system remains robust in such out-of-domain conditions and achieves the best intelligibility.

\begin{table}[ht]
\centering
\caption{Objective evaluation on \texttt{VoxLingua-dev}. 
WER = Word Error Rate, CER = Character Error Rate, estimated by \texttt{Whisper-medium}, 
SPK-SIM is cosine similarity over \texttt{ReDimNet-B6} \cite{yakovlev24_interspeech} embeddings. RTF on an NVIDIA A10 GPU.}
\label{tab:objective}
\begin{tabular}{lcccc}
\toprule
\textbf{System} & \textbf{WER ↓} & \textbf{CER ↓} & \textbf{SPK-SIM ↑} & \textbf{RTF ↓} \\
\midrule
PFluxTTS   & \bf{6.9} & \bf{4.5} & \bf{0.68} & 0.56 $\pm$ 0.02 \\
ChatterBox & 9.0 & 5.9 & 0.61 & 0.54 $\pm$ 0.01 \\
FishSpeech & 45.4 & 35.0 & 0.49 & -- \\
F5-TTS     & 60.2 & 52.7 & 0.58 & 0.25 $\pm$ 0.05 \\
SparkTTS   & 82.5 & 78.0 & 0.23 & 0.28 $\pm$ 0.12 \\
\bottomrule
\end{tabular}
\end{table}

\begin{figure}[t]
    \centering
    \includegraphics[width=0.9\linewidth]{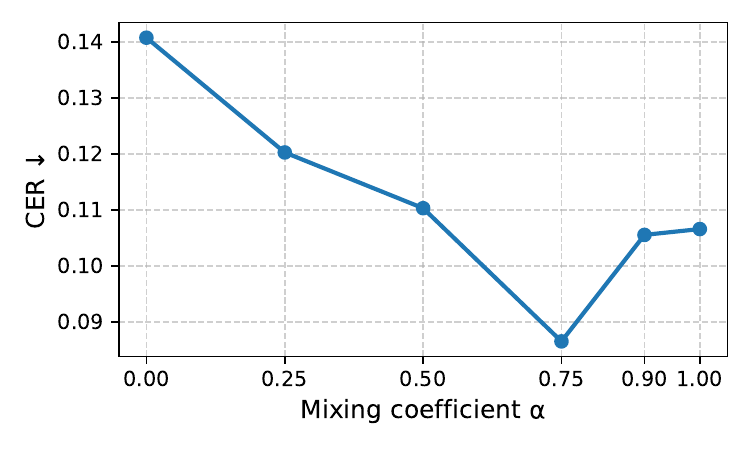}
    \caption{Effect of inference-time model fusion on intelligibility of PFluxTTS (CER as a function of mixing coefficient $\alpha$).}
    \label{fig:fusion_cer}
\end{figure}

\noindent\textit{Ablation studies.}  
We first demonstrate that inference-time model fusion improves intelligibility compared to using either model alone. We computed objective metrics on VoxLingua-dev paired with texts challenging for TTS, following ELLA-V-hard set \cite{song2025ella}.   As shown in Fig.~\ref{fig:fusion_cer}, the AF path alone ($\alpha{=}0.0$) yields a CER of 14.1\%, while the fused model ($\alpha{=}0.75$) reduces CER to 8.6\%. The fused model also outperforms the DG path alone ($\alpha{=}1.0$; CER = 10.6\%).  

Next, we compare the DG-only model with the fused model and show that
adding the AF path improves naturalness.
In a CMOS listening test on 24 samples (based on mTEDx-test) with 10 raters per sample,
the fused model was preferred with a mean $\Delta$CMOS of $0.33$, which is statistically significant
(Wilcoxon signed-rank test, $p<0.012$), with the fused model winning in
79\% of cases.  

Next, we isolate the effect of our FLUX-based prompt conditioning, which uses a sequence of speech prompt embeddings attended to inside the DG decoder, and compare it to a baseline with a fixed prompt embedding injected via AdaLN. Conditioning on a sequence of tokens yields substantially higher speaker similarity: a CMOS test shows a large improvement ($\Delta$CMOS = 1.19 in favor of our method, $p{<}0.05$), and the objective SPK-SIM score increases from 0.47 (fixed embedding) to 0.57 (sequence conditioning).

Finally, we evaluate our PeriodWave+SR vocoder as a standalone waveform generator (Tab.~\ref{tab:vocoder}). We compare against NVSR~\cite{liu2022neural} and BigVGANv2 combined with the pretrained AudioSR model~\cite{liu2024audiosr}. 
Following~\cite{liu2022neural}, we report log-spectral distance (LSD). Our vocoder achieves the best LSD on both datasets, slightly outperforming NVSR on the in-domain \texttt{VCTK-test}, and showing a larger margin on the out-of-domain \texttt{mTEDx} over both baselines.
\begin{table}[t]
\centering
\caption{Log-spectral distance (LSD) evaluated on VCTK-test and mTEDx-test for our PeriodWave-SR vocoder.}
\label{tab:vocoder}
\begin{tabular}{lcc}
\toprule
Method & VCTK-test & mTEDx \\
\midrule
Proposed & \textbf{0.66} & \textbf{1.01} \\
NVSR \cite{liu2022neural} & 0.70 & 1.63 \\
BigVGAN+AudioSR \cite{liu2024audiosr} & 0.99 & 1.39 \\
\bottomrule
\end{tabular}
\end{table}

\section{Conclusion}
We presented PFluxTTS, a system that combines a  duration-guided flow-matching model with an alignment-free model through  inference-time vector field fusion. The duration-guided model is conditioned on a sequence of speech prompts using a FLUX-based approach. The system is paired with a modified  PeriodWave vocoder that performs super-resolution to 48\,kHz with speaker conditioning.

Our experiments focus on English as the target language with multilingual prompts, using real-world conversational datasets such as VoxLingua-dev and mTEDx. We compare our system to strong baselines and confirm the contributions of model fusion, FLUX-based conditioning, and the super-resolution vocoder via targeted evaluations. Overall, PFluxTTS outperforms most open-source systems and a leading commercial solution, achieving subjective results comparable to ChatterBox (with lower WER), despite ChatterBox being trained on nearly an order of magnitude more data.

In the future, we plan to scale training to larger datasets, investigate alternative fusion schedules, and develop methods to control the balance between prompt-driven and text-driven prosodic features.

\bibliographystyle{IEEEbib}
\bibliography{refs}

\end{document}